\documentclass[journal]{IEEEtran}
\usepackage[dvips]{graphicx}
\usepackage[fleqn]{amsmath}
\usepackage{amsfonts}
\usepackage{amssymb}
\usepackage{amsthm}
\usepackage{amsopn}

\ifCLASSINFOpdf
  
\else
  
\fi

\begin{document}

\title{ECG beats classification using waveform similarity and RR interval}

\author{Ahmad Khoureich Ka
\thanks{A.~K.~Ka: Institut de Recherche Math\'{e}matique de Rennes. Universit\'{e} de Rennes 1 - Campus de Beaulieu 35042 Rennes Cedex - France.  Universit\'{e} Cheikh Anta Diop de Dakar. e-mail: khoureich@yahoo.fr.}% <-this % stops a space
}

%\markboth{Journal of \LaTeX\ Class Files,~Vol.~6, No.~1, January~2007}%
%{Shell \MakeLowercase{\textit{et al.}}: Bare Demo of IEEEtran.cls for Journals}
\maketitle

\begin{abstract}
This paper present an electrocardiogram (ECG) beat classification method based on waveform similarity and RR interval. The purpose of the method is to classify six types of heart beats (normal beat, atrial premature beat, paced beat, premature ventricular beat, left bundle branch block beat and right bundle branch block beat). The electrocardiogram signal is first denoised using wavelet transform based techniques. Heart beats of 128 samples data centered on the R peak are extracted from the ECG signal and thence reduced to 16 samples data to constitute a feature. RR intervals surrounding the beat are also exploited as feature. A database of annotated beats is built for the classifier for waveform comparison to unknown beats. Tested on 46 records in the MIT/BIH arrhythmia database, the method shows classification rate of 97.52\%.
\end{abstract}


\IEEEpeerreviewmaketitle

\section{Introduction}

\IEEEPARstart{W}{ith} the introduction of the string galvanometer by Willem Einthoven the electrocardiogram (ECG) has become one of the most important tools in the diagnosis of heart deseases. The ECG is the graphical display of electrical activity of the heart recorded from electrodes on the body surface~\cite{Fra}.
From the plot of an ECG a cardiologist can analyse the shape of the waveform and determine the nature of deseases afficting the heart. The abnormal beats in the ECG pointing to a particuliar desease can be rare and widespread in the span of a large record. Therefore, the work of the cardiologist tracking down abnormalities  can be tedious. Thus it becomes very helpful to use computer-based diagnosis. 

Besides the fact the ECG record can be noisy, the main problem in computer based classification is the wide variety in the shape of beats belonging to the same class and beats of similar shape belonging to differents classes~\cite{Oso,Lia}. Therefore the algorithms in computer-based diagnosis are generaly of three steps: EGC beat detection, extraction of usefull features from beats and classification. For beat detection a number of methods are available in the literature~\cite{Afo,Koh}. Feature extraction can be done in time domain~\cite{Cha}, in frequency domain~\cite{Ach}, by multiscale decomposition~\cite{Pra}, by multifractal analysis~\cite{Iva} or by statistical means~\cite{Oso}. The classification can be performed by neural networks~\cite{Ozb,Mar}, mixture of experts~\cite{HuY}, switchable scheme~\cite{Sun}.

Although statistical methods of ECG beats recognition~\cite{Oso,Sun,Meh}, reports good recognition accuracy, we believe as in the work of Yu Hen Hu~\cite{HuY} that a good approach in ECG beat classification is to take into account specificities of each patient's electrocardiogram. Therefore the method proposed in this paper is a patient-specific classifier. Wavalet transform has been successfully used in the processing of non stationary signals like electrocardiograms~\cite{Lot}. Therefore it is employed in this study for ECG signal denoising and beat length reduction. A beats database containing five classes of annotated beats is created for the classifier. And each time a patient's ECG beats have to be classified the five first minutes manually annotated beats of the ECG record is integrated into that beats database. So the classifier experience grows up each time an ECG record is submited to it for automatic annotation. The beats of the patient's ECG are first clustered by similarity in the shape of their waveform and then each cluster is classified by considering the greater similarity of its elements to the beats in the classifier beats database.

\section{ECG signal pre-processing}
The purpose of this work is to classify six types of heart beats which are : Normal beats (N), Premature Ventricular Contractions (PVC), Paced Beats (PB), Atrial Premature Beat (APB), Left Bundle Branch Block beats (LBBB) and Right Bundle Branch Block beats(RBBB). A description of these arrhythmias can be found at~\cite{Fra}. Forty six (46) ECG signals recorded with the Mason - Likar II lead (MLII) are taken from the MIT/BIH arrhythmia database for the creation of the beats database and the evaluation of the classifier.

The ECG records are generaly noisy, they present a baseline wander and high frequency noise. Techniques based on discrete wavelet transform are used to overcome these disturbances.

\subsection{Discrete Wavelet Transform}
The Discrete Wavelet Transform is mainly based on the multiresolution analysis of the wavelet transform introduced by S. Mallat~\cite{Mal}. With the discrete wavelet transform any function $f \in L_2(\mathbb{R})$ can be uniquely represented in terms of an $L_2$-convergent series:
\begin{equation}
 x(t) = \sum_k \alpha_{j_0k}\phi_{j_0k}(t) + \sum_{j=j_0}^{\infty}\sum_k \beta_{jk}\psi_{jk}(t), 
\end{equation}
where $\{\phi_{j_0k}\}$ is an orthonormal system from the scaling function and $\{\psi_{jk}\}$ an orthonormal system from the mother wavelet, and 
$$ \alpha_{j_0k} = \int x(t)\overline{\phi_{j_0k}(t)}dt, \qquad \beta_{jk} = \int x(t)\overline{\psi_{jk}(t)}dt,$$ are the wavelet coefficients. The cascade algorithm~\cite{Mal} allows the computation of the lower level coefficients from the higher level ones and vice versa:
\begin{equation}
 \alpha_{jk} = \sum_{l}h_{l-2k}\alpha_{j+1,l} \quad \textmd{and} \quad \beta_{jk} = \sum_{l}\lambda_{l-2k}\alpha_{j+1,l}
\end{equation}
where $\lambda_{k} = (-1)^{k+1}h_{1-k}$, $j$ and $k$ are integers and $\{h_k\}$ are the mother wavelet coefficients. The index $j$ indicates the resolution level of the multiresolution analysis. The inverse transformation is given by the following equation:
\begin{equation}
 \alpha_{j+1,l} = \sum_{k}h_{l-2k}\alpha_{jk} + \sum_{k}\lambda_{l-2k}\beta_{jk}
\end{equation}

\subsection{ECG signal noise reduction techniques}
ECG records are generaly corrupted by noise from different sources. The noise can appear as a baseline wander and/or a high frequency oscillation along the signal.

In this work, for baseline wander cancellation we have used the method proposed in~\cite{Ant} for its ability to eliminate baseline drift without including distortion in the signal. On the other hand we have delt with the high oscillation noise by using the soft thresholding technique proposed by D. Donoho~\cite{Don}.

\begin{figure}[!h]
\centering
\includegraphics[angle=270,width=88mm]{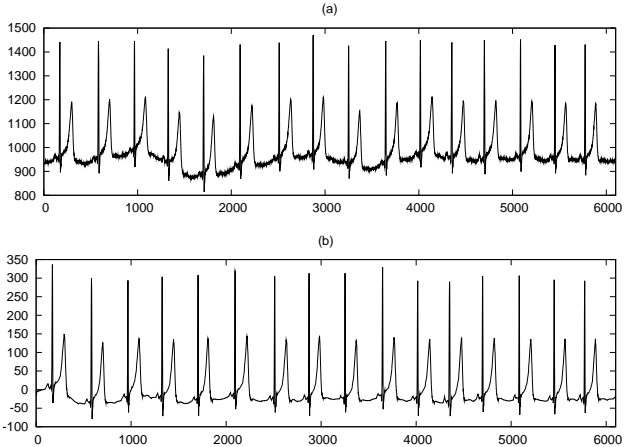}
\caption{Noise cancellation result on record number 113, MLII, from MIT-Arrhythmia Database: (a) Original signal (b) Denoised signal.}
\label{fig_ecg}
\end{figure}

We have made a java implementation of these two noise reduction techniques and run the program on a noisy ECG signal. The result is shown in figure~\ref{fig_ecg}.

\section{Proposed method}

\subsection{Feature extraction}
It is known that the morphology of the QRS complex along with the instantanuous RR interval (interval between two successives R peaks) play an important role in the heart deseases diagnosis~\cite{Sun,Meh}. For the classification of an ECG beat the ratio of the RR interval before it to the one after it is a usefull feature. For beats such as N, LBBB, RBBB and PB this ratio is near or equal to 1 and for beats in the classe of APB and PVC this ratio is rather less than 1. But sometimes it happens that PVCs or APBs occur in groups~\cite{Fra}, in such case the ratio can be near or equal to 1 but the RR interval will be very small compared to the mean RR of none premature beats in the ECG signal. The distinction between PVC and APB is that usually a PVC is followed by a compensatory pause i.e., the RR interval between two QRS enclosing PVC equals twice the normal RR interval; in contrast, APBs are usually followed by no compensatory pause i.e., the RR interval between two QRS enclosing APB is less than twice the normal RR interval~\cite{Fra}.

Finally $RR_b, RR_a$ (respectively the RR interval before and after the R peak) along with $RR_m$ the mean RR of no premature beats are considered as features in this study.

The R peaks are taken from the annotation files of the MIT-BIH arrhythmia database. From the ECG record, beats are extracted using 128 samples centered at the R peak. After that, the beat size is reduced to 16 samples by discrete wavelet transform as follows:
with the multiresolution approch the 128 samples beats at resolution level $j_0$ is denoted by
\begin{equation}
 b_{128}(t) = \sum_{k=0}^{2^{j_0}-1}\alpha_{j_0k}\phi_{j_0k}(t),\qquad \beta_{j_0k} = 0,
\end{equation}
its decomposition to a resolution level $j_0-3$ is 
\begin{equation}
 b_{128}(t) = \sum_{k=0}^{2^{j_0-3}-1}\alpha_{j_0-3,k}\phi_{j_0-3,k}(t) + \sum_{j=j_0-3}^{j_0}\sum_{k=0}^{2^j-1}\beta_{jk}\psi_{jk}(t),
\end{equation}
$\phi$ and $\psi$ are scale functions and wavelet functions at the corresponding resolution level. 

The beats at resolution level $j_0$ can be characterized by $2^{j_0}=128$ samples per length unit, then $j_0=7$, thus the last equation becomes
\begin{equation}
  b_{128}(t) = \sum_{k=0}^{2^4-1}\alpha_{4,k}\phi_{4,k}(t) + \sum_{j=4}^7\sum_{k=0}^{2^j-1}\beta_{jk}\psi_{jk}(t)
\end{equation}

By setting all the detail  coefficients $\beta_{jk}$ to zero we consider the reduced beat 
\begin{equation}
  b(t) = \sum_{k=0}^{2^4-1}\alpha_{4,k}\phi_{4,k}(t)
\label{equ:bt}
\end{equation}
as a feature in our method. The reduction of the beat size from 128 to 16 samples saves hardware memory and accelerates the processing speed.

\subsection{Classification}
Let $\mathbb{A}$ be the orthonormal family of scaling functions $\{\phi_{4,k}\}$ at resolution level $2^4$. From equation (\ref{equ:bt}) we see that $\mathbb{A}$ generates a vector space $\mathbb{B} \subset L_2(\mathbb{R})$ containing all the beats in our study. Therefore the scalar product in $L_2(\mathbb{R})$ can also be defined for every two vectors of $\mathbb{B}$~:
$\vert\Psi\rangle=\sum_{\phi\in\mathbb{A}}\Psi_\phi\vert\phi\rangle$\quad and\quad $\vert\Psi^\prime\rangle=\sum_{\phi\in\mathbb{A}}\Psi^\prime_\phi\vert\phi\rangle$\quad by\quad  $\langle\Psi\vert\Psi^\prime\rangle=\sum_{\phi\in\mathbb{A}}\overline{\Psi}_\phi\Psi^\prime_\phi$, where $\overline{\Psi}_\phi$ denote the complex conjugate of $\Psi_\phi$ (it coincides with $\Psi_\phi$ if it is real).
We can define a similarity function on $\mathbb{B}$ as follow:
\begin{equation}
 \sigma\colon\mathbb{B}\times\mathbb{B}\rightarrow\lbrack0,1\rbrack, \qquad \sigma(\Psi,\Psi^\prime)=\frac{\langle\Psi\vert\Psi^\prime\rangle}{\Vert\Psi\Vert\cdot\Vert\Psi^\prime\Vert}
\end{equation}
$\Vert\cdot\Vert$ denotes a Hilbert norm on $\mathbb{B}$. Applied on two beats the similarity function indicates the degree of proximity in their shape. Such similarity functions have been used previously in literature in many different contexts, see~\cite{Tho} for a recent up to date review.

As stated in~\cite{AH1} we believe that computer analysis cannot substitute physician's interpretation of ECG. Therefore our classifier use a database of known beats (reduced length version of the original 128 samples beats) taken within the five first minutes of each record in the MIT/BIH arrhythmia database. The classifier beat database contains five classes of beats which are N, LBBB, RBBB, PB and PVC. The APB beat type is not present in that database because APBs can be similar to N, LBBB and RBBB, therefore its identification is mainly based on the ratio of the previous RR interval to the following one ($RR_b/RR_a$). The beats database acts as the classifier knowledge. It grows up each time a patient's ECG record is treated because the manually annotated beats in the five first minutes are included in it. This practice is conformed to the AAMI (American Association of Medical Instrumentation) recommended procedure which allows the use of the first 5 minutes of data in an ECG record to fine tune the classifier~\cite{HuY}.

After the denoising step, the classification of a patient's ECG record is done with the following steps:

(1) Beats are extracted using 128 samples data centered on the R peaks. For each beat the RR interval before ($RR_b$) and after ($RR_a$) its R peak are taken. Afterwards beat length are reduced to 16 samples using discrete wavelet decomposition as described in the feature extraction section. The resulted beat is normalized to reduce waveform differencies in the same class by subtracting the mean value and then dividing by the standard deviation~\cite{Sun}.

(2) The beats within the 5 first minutes of the ECG are included in the classifier beat database. And the mean RR ($RR_m$) of beats in those 5 minutes is taken as feature.

(3) The beats in the 25 minutes remaining are first hierarchical clustered by similatity before classification.

(4) The class of a beat is identified using its highest similarity to beats in the classifier database.

(5) If the  class is found to be N or LBBB or RBBB, the ratio $RR_b/RR_a$ is calculated.

If $RR_b/RR_a < (1-\epsilon_1)$ ou $(RR_a + RR_b) < (2RR_m - \epsilon_2)$ then the class is changed to APB otherwise the class is unchanged. The optimal $\epsilon_1$ and $\epsilon_2$ value will be identified with experiments.

(6) The class of a cluster is the class of its element which has the highest similarity to a beat in the classifier ECG database.

\section{Results and discussion}
The classifier was tested with the 25 minutes of data from each 30 minutes ECG signals recorded with the Mason - Likar II lead (MLII) in the MIT/BIH arrhythmia database. The 5 first minutes of those ECG signals were used to build the beats database used by the classifier. In each ECG record used in this study only beats in the class of normal beat (N), left bundle branch block beat (LBBB), right bundle branch block beat (RBBB), artial premature beat (APB), premature ventricular contraction (PVC) and  paced beat (PB) are considered. The RR intervals are calculated from the position of the R peak documented in the annotation files of the MIT/BIH database. The mean RR ($RR_m$) used in the identification of atrial premature beats (APB) is calculated using the 5 first minutes of the signal. For the clustering of beats before classification, we consider that two beats belong to the same cluster if their similarity is greater or equals to 0.95.
\begin{table}[!ht]
\label{tab1}
\caption{Classification results}
\centering
\begin{tabular}{|c|c|c|c|}
\hline
 Record \# &  \# Beat &  \# Misclassification &  Classif. rate (\%)\\\hline
100 & 1892 & 0 & 100\\
\hline
101 & 1514 & 2 & 99.87\\
\hline
103 & 1721 & 5 & 99.71 \\
\hline
105 & 2141& 71& 96.68 \\
\hline
106 & 1688& 52& 96.92 \\
\hline
107 & 1776& 11& 99.38\\
\hline
108 & 1468& 53& 96.39\\
\hline
109 & 2089& 11&99.47 \\
\hline
111 & 1768& 41&97.68 \\
\hline
112 &2101 & 2& 99.90\\
\hline
113 & 1498& 3& 99.80\\
\hline
114 & 1590& 44& 97.23\\
\hline
115 & 1628& 1& 99.94\\
\hline
116 & 2007& 23& 98.85\\
\hline
117 & 1277& 3& 99.77\\
\hline
118 & 1907& 42& 97.80\\
\hline
119 & 1653& 0& 100\\
\hline
121 & 1551& 5& 99.68\\
\hline
122 & 2044& 0& 100\\
\hline
123 & 1262& 0& 100\\
\hline
124 & 1334& 12& 99.10\\
\hline
200 & 2156& 76& 96.47\\
\hline
201 & 1408& 44& 96.88\\
\hline
202 & 1860& 63& 96.61\\
\hline
203 & 2467& 592& 76\\
\hline
205 & 2180& 6& 99.72\\
\hline
207 & 1475& 21& 98.58\\
\hline
208 & 2122& 19& 99.10\\
\hline
209 & 2509& 109& 95.66\\
\hline
210 & 2164& 57& 97.37\\
\hline
212 & 2275& 7& 99.69\\
\hline
213 & 2450& 40& 98.35\\
\hline
214 & 1866& 86& 95.39\\
\hline
215 & 2780& 104& 96.26\\
\hline
217 & 1608& 12& 99.25\\
\hline
219 & 1764& 44& 97.51\\
\hline
220 & 1685& 36& 97.86\\
\hline
221 & 2011& 2& 99.90\\
\hline
222 & 1892& 279& 85.25\\
\hline
223 & 2166& 265& 87.77\\
\hline
228 & 1695& 29& 98.29\\
\hline
230 & 1849& 2& 99.89\\
\hline
231 & 1270& 12& 99.06\\
\hline
232 & 1478& 16& 98.92\\
\hline
233 & 2543& 44& 98.27\\
\hline
234 & 2231& 0& 100\\
\hline
Average & & & 97.52\\\hline
\end{tabular}
\end{table}

The performance of the classifier is evaluated in terms of accuracy rate  by ECG signal and overall accuracy. The classification results are summarized in Table 1.

The overall classification result is found to be 97.52\%. This is a good recognition accuracy in regard to the fact exposed in~\cite{AH2} that the percentage of ECGs correctly classified by the computer programs have a median of 91.3\%.

In some records (203, 222 and 223) the classification rate is rather low. An examination of these records indicates a high variation of length of the RR intervals around normal beats. This variations causes normal beats to be classified as APBs. Another fact to take in consideration is that misclassification can result from errors in the position of R peaks provided by the manually annotated files in the MIT-BIH database. Our similarity function can give bad results for two beats in the same class when their R peak are misaligned at the top of their QRS complex. We are currently developping a promising method for the detection of the summit of R peaks for better alignment when comparing two beats.

A little difficulty in the application of our method can be the creation of the beats database used by the classifier. We think, it is possible to find a way to reduce the 5 minutes a cardiologist have to annotate. But we stay convinced as in~\cite{AH2} that  Computer-based interpretation of the ECG is an adjunct to the electrocardiographer, and all computer-based reports require physician overreading.

A comparison of our method to other works in the field of automatic ECG beat classification is summerized in table 2. But actually, efficency comparison of methods is not straighforward due to the difference in the test conditions (type of beats to classify and number of beats used for test).
\begin{table}[!ht]
\caption{Comparative results of different ECG beat classification methods}
\centering
\begin{tabular}{|l|c|c|}
\hline
Method & Number of beat type & Accuracy(\%)\\\hline
ICA~\cite{Sun} & 6 & 99.51\\\hline
FTNN~\cite{Min} & 3 & 98\\\hline
MOE~\cite{HuY} & 4 & 94\\\hline
MRANN~\cite{Pra} & 13 & 96.79\\\hline
FHNN~\cite{Oso} & 7 & 96.6\\\hline
Our method & 6 & 97.52\\\hline
\end{tabular}
\label{tab2}
\end{table}

It is interesting to note that some methods published in the literature are not tested on a large number of beats as we do. For exemple in the Independent Component Analysis~\cite{Sun} method, authors have used per ECG signal 100 beats for training and 100 beats for testing even if the signal contains more than 2000 beats. Since the morphology of beats of the same type not only changes from patient to patient but also within the same patient~\cite{Lia}, the number of beats for testing can impact on the recognition accuracy. In the table 1 if we get rid of the records 203, 222 and 223 which give bad results the overall classification rate of our method increases from 97.52 to 98.53\%.

\section{Conclusion}
In this paper, we present a patient-adaptable ECG beat classification method based on a similarity function and a beats database which acts as the classifier knowlegde. Discrete wavelet transform is also used for the ECG signal preprocessing. The method uses a simple approach and runs with low processing cost in comparison with those using neural networks or fuzzy logic. A promising accuracy in the classification of six types of heart beats has been reached.

\end{document}